\documentclass{jps-cp}
\usepackage{txfonts} 

\usepackage{subfigure}
\usepackage{color}

\title{Development of a UV-transparent Lens Array for Enlarging the Effective Area of Multichannel SiPMs}

\author{Akira~\textsc{Okumura}, Akira~\textsc{Asano}, Kazuhiro~\textsc{Furuta}, Naoya~\textsc{Hidaka}, Yuki~\textsc{Nakamura}, Hiroyasu~\textsc{Tajima}, and Anatolii~\textsc{Zenin}}

\inst{Institute for Space--Earth Environmental Research, Nagoya University, Furo-cho, Chikusa-ku, Nagoya 464-8601, Japan}

\email{oxon@mac.com (A.O.)}

\recdate{June 5, 2019}

\abst{We developed a UV-transparent lens array that can increase the photon detection efficiency of a silicon photomultiplier (SiPM) array comprising of 64 pixels ($3\times3$~mm$^2$ each) and 0.2-mm gaps. Through the plano-convex spherical lens on each $3.2\times3.2$~mm$^2$ region, we showed that the loss of photon detection efficiency due to the pixel gaps could be recovered as the incident photons get concentrated on the sensitive regions of the SiPM array. By using a prototype lens array, we achieved approximately 10\%--30\% relative increase in photon detection efficiency in our target angles of incidence of 30--60~deg.}

\kword{Silicon photomultipliers (SiPMs), lens array, ray-tracing simulation, Cherenkov telescopes}

\begin{document}
\maketitle

\section{Introduction}

Silicon photomultipliers (SiPMs) are widely used in various physics experiments owing to their compactness, low operation voltage, and high photon detection efficiency (PDE). PDE is defined as a product of quantum efficiency (QE) and other factors, namely, the geometrical fill factor and probability of the avalanche process of individual Geiger-mode avalanche-photodiode (G-APD) cells for SiPMs and the collection efficiency of 90\%--95\% for conventional photomultiplier tubes (PMTs).

SiPM PDE reaches as high as 50\% when a G-APD cell size of 75-$\mu$m is chosen (e.g., Hamamatsu Photonics S13360 series with a fill factor of 82\%), whereas that of PMTs with an ultrabialkali photocathode (maximum QE of typically nearly 43\%) can achieve approximately 40\% at most. Thus, the use of SiPMs could be a more preferable option than PMTs in low photon luminosity and compact multichannel applications, such as small focal-plane ($<50$~cm) cameras of ground-based very-high-energy (100~GeV--10~TeV) gamma-ray telescopes (also referred to as Cherenkov telescopes).

\begin{figure}[tbp]
  \centering
  \includegraphics[clip, width=12cm]{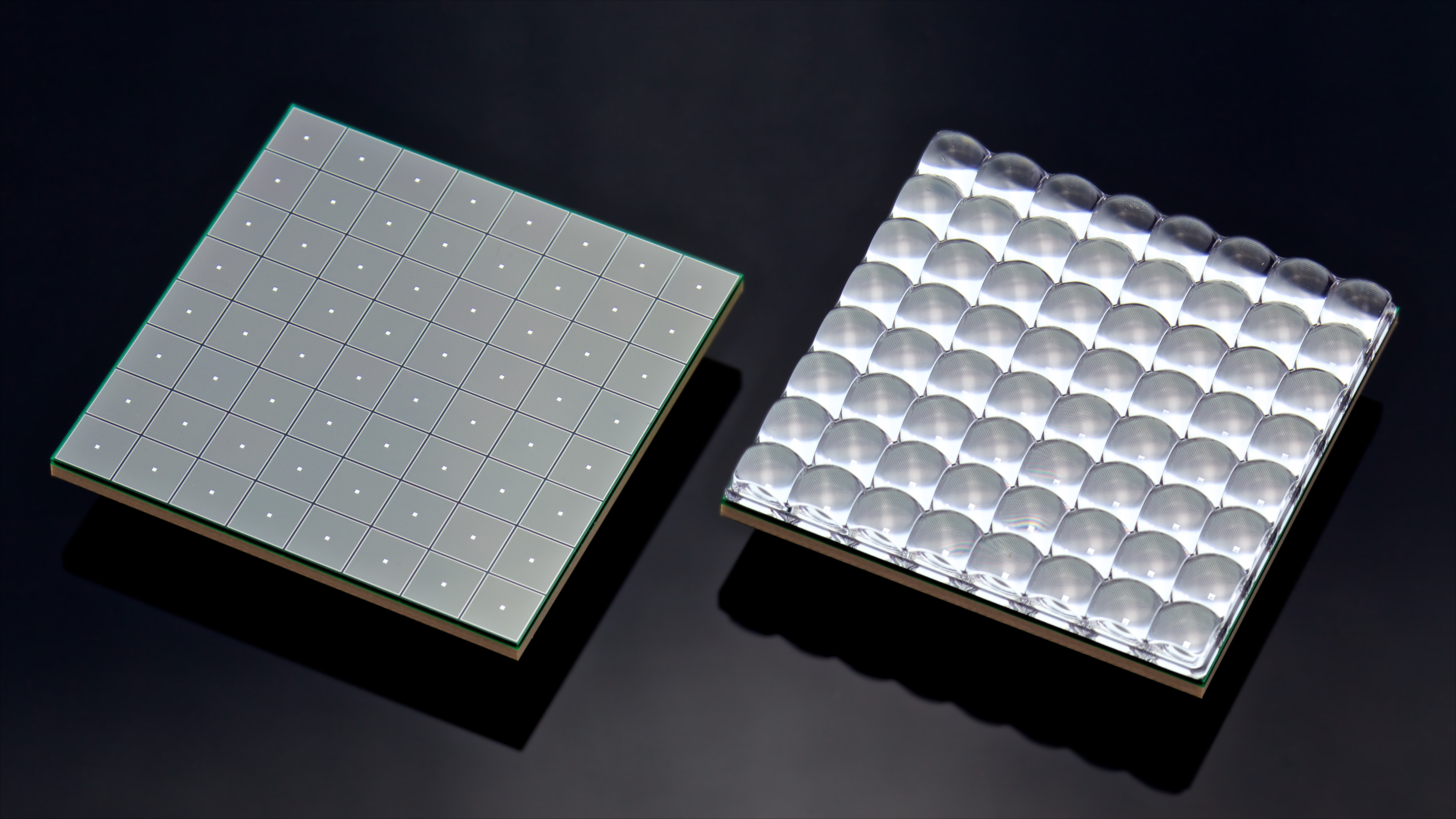}
  \caption{$8\times8$ pixel SiPM (Hamamatsu Photonics S13361-3050AS) (left) and the same product with a lens array (right).}
  \label{fig:lens_array}
\end{figure}

\begin{figure}[tbp]
  \centering
  \subfigure[]{%
    \label{fig:pixel-wo-lens}
    \includegraphics[clip, width=6cm]{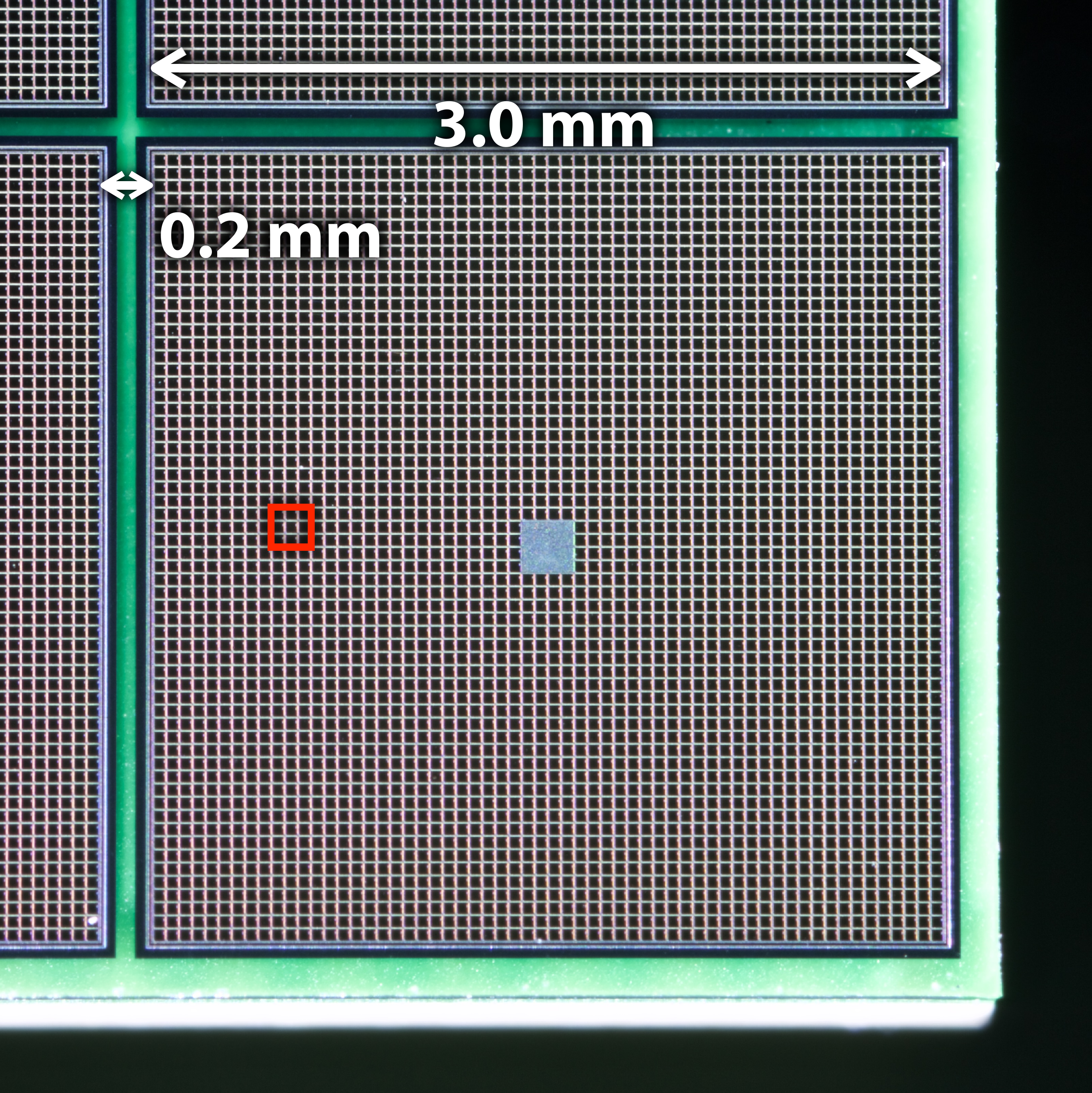}
  }
  \subfigure[]{%
    \label{fig:pixel-with-lens}
    \includegraphics[clip, width=6cm]{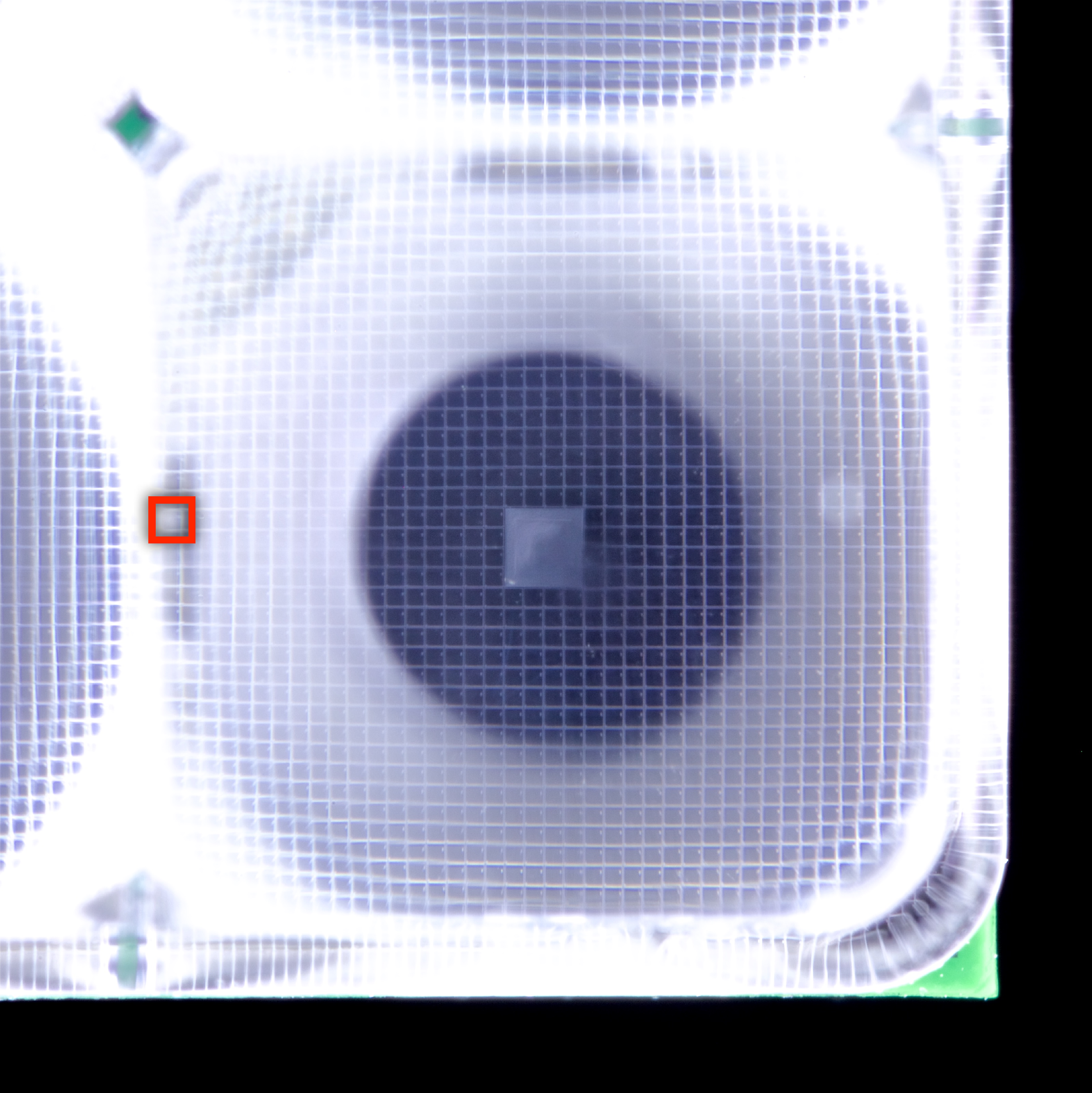}
  }
  \caption{(a) Close-up view of one of the corner pixels of the SiPM array, and 3584 G-APD cells and the central metal pad are presented. (b) The same pixel with a lens array. The G-APD cells and the pad are apparently magnified because of the convex lens shape. The circular central shadow with a radius of approximately 10 G-APD cells is a reflected image of a camera lens. Small red squares in both photos point the identical G-APD cell.}
\end{figure}

Covering the entire focal-plane area with a large number of SiPMs with the smallest possible dead area is ideally desirable. However, non-zero gaps between two adjacent channels usually exist in multichannel SiPM products (e.g., $0.2$ mm for the Hamamatsu Photonics S13361 series), as shown in Figs.~\ref{fig:lens_array} and \ref{fig:pixel-wo-lens}, thereby resulting in a non-negligible effective PDE loss amounting to relatively $12.1$\% ($=1-3.0^2/3.2^2$) for $3\times3$-mm$^2$ pixels with 0.2-mm gaps. In relation to one of the advantages of SiPMs, the high PDE becomes less significant, unless this PDE loss is recovered using some techniques.

On the contrary, intentionally having wider pixel gaps can reduce the production cost of a large SiPM array, as the SiPM unit price under mass production is roughly proportional to the total area of the silicon substrate. On such basis, having a higher effective PDE and building a less expensive SiPM array come as trade-offs.

Installing solid light concentrators called compound parabolic concentrators (CPCs, also known as Winston cones) on individual single-channel SiPMs is a feasible method for identifying a less effective PDE loss, considering a smaller silicon area \cite{cite:FACT}. However, the absorption of ultraviolet (UV: 300--400~nm) photon inside long solid cones (a few cm) is not negligible; thus, it is not extremely suitable for UV--visible Cherenkov photon detectors. The use of hollow Winston cones with an optimized Bézier curve profile has been proposed and demonstrated \cite{cite:Okumura-Cone,cite:SST-1M}. However, the minimum plastic thickness required in injection molding is approximately 0.2~mm, a size that produces an unavoidable pixel gap of 0.4~mm in total.

In the present study, we explored another solution to reduce the effective PDE loss caused by the pixel gaps by using a lens-array molding technology and a UV-transparent glass material. A discussion of lens array design, simulation, and measurements and a comparison of the effective PDE between SiPMs with and without a lens array were carried out.

\section{Lens-array Design}

A micro lens array (roughly 10~$\mu$m) on an imaging sensor is quite commonly used in commercial digital camera products with complementary metal oxide semiconductor (CMOS) or charge-coupled device (CCD) imaging sensors with a typical pixel size of $5$--$10$~$\mu$m. Convex micro lenses aligned on individual imaging pixels concentrate incident photons onto the active area of the pixels to achieve higher sensitivity.

Accordingly, it is natural to apply the same idea to multichannel SiPMs to achieve a higher effective PDE with the silicon substrate area being kept small. Figures~\ref{fig:pixel-wo-lens} and \ref{fig:pixel-with-lens} clearly show that the lens-array concept can optically hide the 0.2-mm gaps between 3-mm pixels, where a lens array comprising of $8\times8$ plano-convex spherical lenses is placed and coupled to the SiPM array with optical grease (Saint-Gobain BC-630, refractive index $n$ of approximately 1.5).

Each lens has a $3.2\times3.2$~mm$^2$ footprint to cover a single SiPM pixel. The lens height and the radius of curvature are 2.0~mm and 2.3~mm, respectively, which are the chosen values to maximize the effective PDE for the Cherenkov photon spectrum at the ground level. Moreover, a flat distribution of angles of incidence from 30 to 60~deg was assumed to consider a future Cherenkov camera application with the Schwarzschild--Couder optical system\cite{cite:SC}. As illustrated in Fig.~\ref{fig:pixel-with-lens}, the radius of curvature is too short for normal incidence photons, and the outer G-APD cells are totally hidden. On the contrary, a large curvature is required to concentrate photons with large angles of incidence of up to 60~deg onto the cells.

\section{Simulation}

During design optimization, we simulated the performance of the lens array assuming the PDE curve of Hamamatsu Photonics S12571-050C (50-$\mu$m cell size, silicone resin coating). The wavelength-dependent photon absorption length, $\tau_\mathrm{abs}(\lambda)$, and refractive indices, $n(\lambda)$, of the silicone resin, optical grease, and UV-transparent lens-array material (OMG UVC-200B) were measured and considered in the simulation. For simplicity, the Fresnel reflection and the interference at the media boundary on the silicon substrate were ignored. The effective PDE increase for a flat distribution of angles of incidence (30--60~deg) for a Schwarzschild--Couder Cherenkov camera under these assumptions, as simulated by the ROBAST ray-tracing software\cite{cite:ROBAST}, was 13.5\%. For example, if the PDE without the lens array was 40\%, then the effective PDE with the lens array would be $40\times1.135=45.4$\%. Consequently, the effective PDE loss of $12.1$\% due to the pixel gaps would be almost compensated by the lens array ($\frac{1}{1-0.121}=1.138$).

We performed a more realistic simulation after the production and performance test of the lens array to reproduce the measurement result. Thus, the G-APD cell structure, thin $\mathrm{SiO}_2$ and $\mathrm{Si}_3\mathrm{N}_4$ layers on the silicon substrate, and $n(\lambda)$ of the layers and the substrate were additionally considered, whereas the optical interference on the layers remained ignored, as ROBAST does not support quantum optics.

\begin{figure}[tbp]
  \centering
  \subfigure[310~nm]{%
    \label{fig:310nm}
    \includegraphics[clip, width=0.48\linewidth]{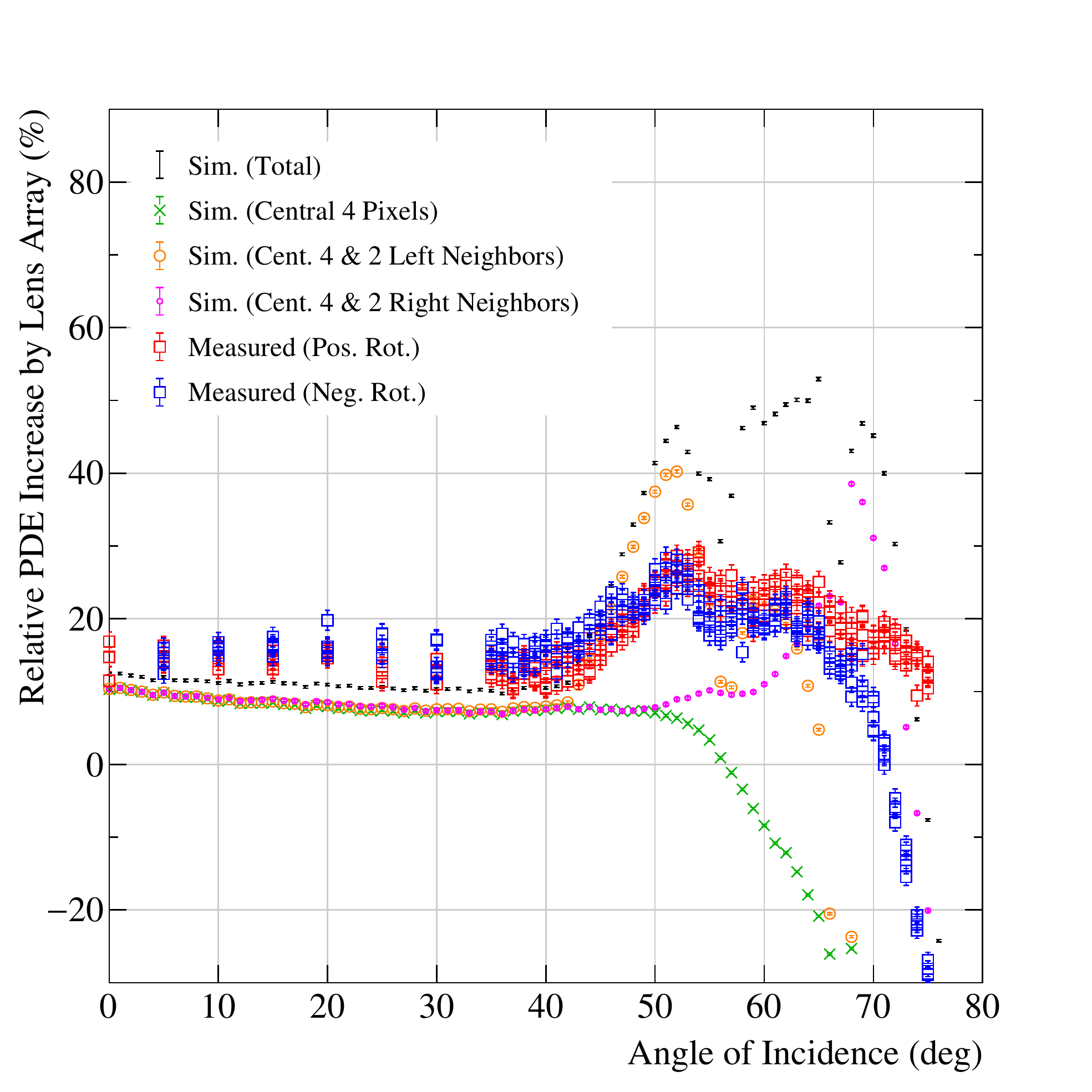}
  }
  \subfigure[375~nm]{%
    \label{fig:375nm}
    \includegraphics[clip, width=0.48\linewidth]{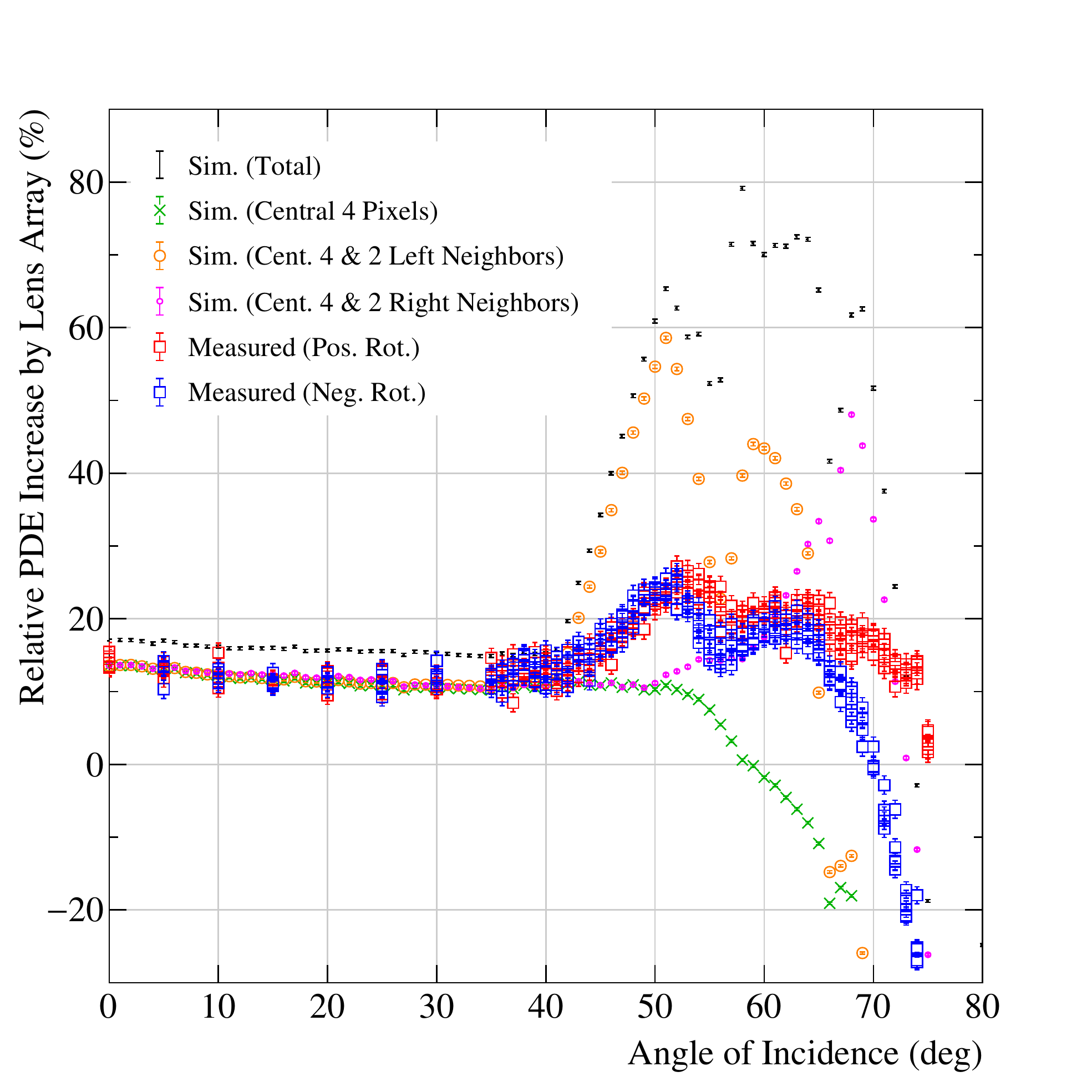}
  }\\
  \subfigure[465~nm]{%
    \label{fig:365nm}
    \includegraphics[clip, width=0.48\linewidth]{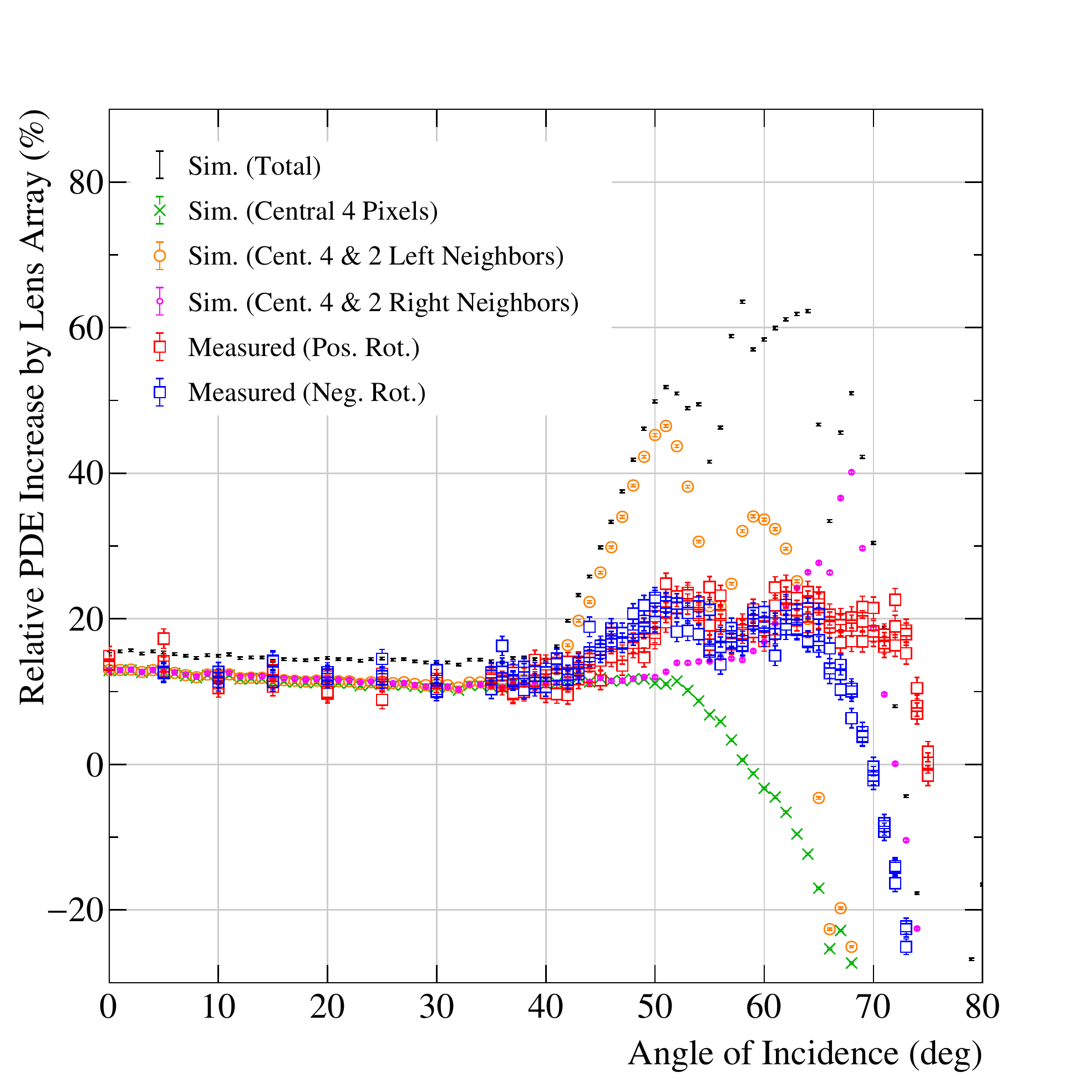}
  }
  \subfigure[635~nm]{%
    \label{fig:635nm}
    \includegraphics[clip, width=0.48\linewidth]{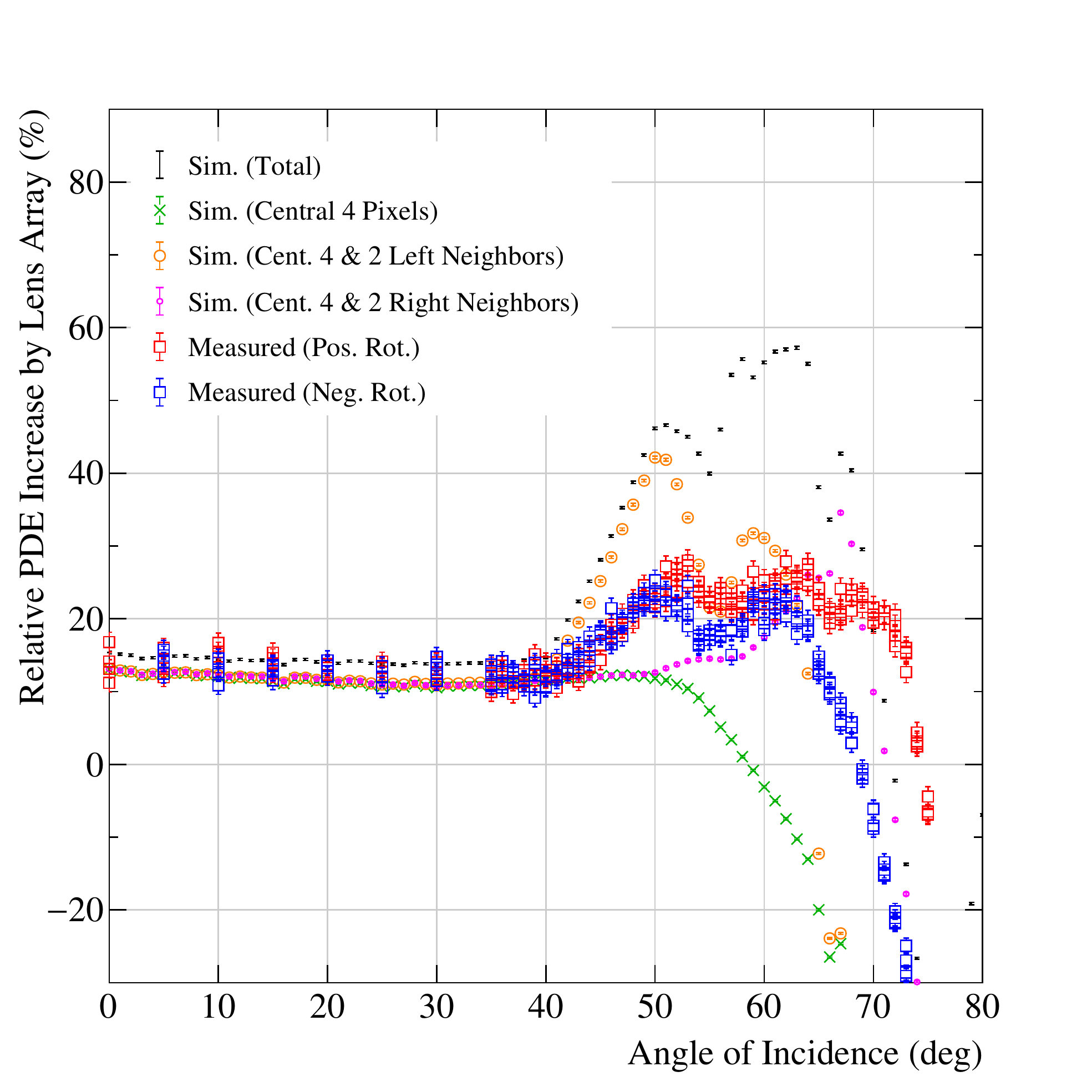}
  }
  \label{fig:sim_meas}
  \caption{Relative PDE increase by lens array vs. angle of incidence for four LED colors. ROBAST simulations for different pixel groupings are shown as total (black), central four pixels only (green crosses), central four pixels and two left neighbor pixels only (orange circles), and central four pixels and two right neighbor pixels only (small magenta circles). See also Fig.~\ref{fig:map_465nm_50deg}. Red and blue open squares show the measured increase for positive and negative rotations, respectively.}
\end{figure}

The ROBAST simulation for different LED colors (310, 375, 465, and 635~nm) are presented in Fig.~\ref{fig:sim_meas}. The measurement results are discussed in Section~\ref{sec:measurement}. Simulated photons were beamed onto the central $3.2\times3.2$-$\mathrm{mm}^2$ region of the $8\times8$ SiPM array from outside of the lens array. The relative PDE increase in 0--40~deg reached nearly 10\%--12\%, 13\%--17\%, 14\%--16\%, and 14\%--15\% for 310, 375, 465, and 635~nm, respectively, where all the photons detected by any SiPM channel were counted (Total, in black color in Fig.~\ref{fig:sim_meas}). With the high refractive index of silicon (approximately 3.9--6.9 in 300--600~nm), a large fraction of photons got reflected on the silicon surface, and a part of them was totally reflected at the media boundary between the lens array and air. Thus, photons with large angles of incidence exhibited a higher probability of reaching the silicon surface multiple times, resulting in higher PDE increase than that in 0--40~deg.

\begin{figure}[tbp]
  \centering
  \subfigure[Count map (50~deg)]{%
    \label{fig:map_465nm_50deg}
    \includegraphics[clip, width=6cm]{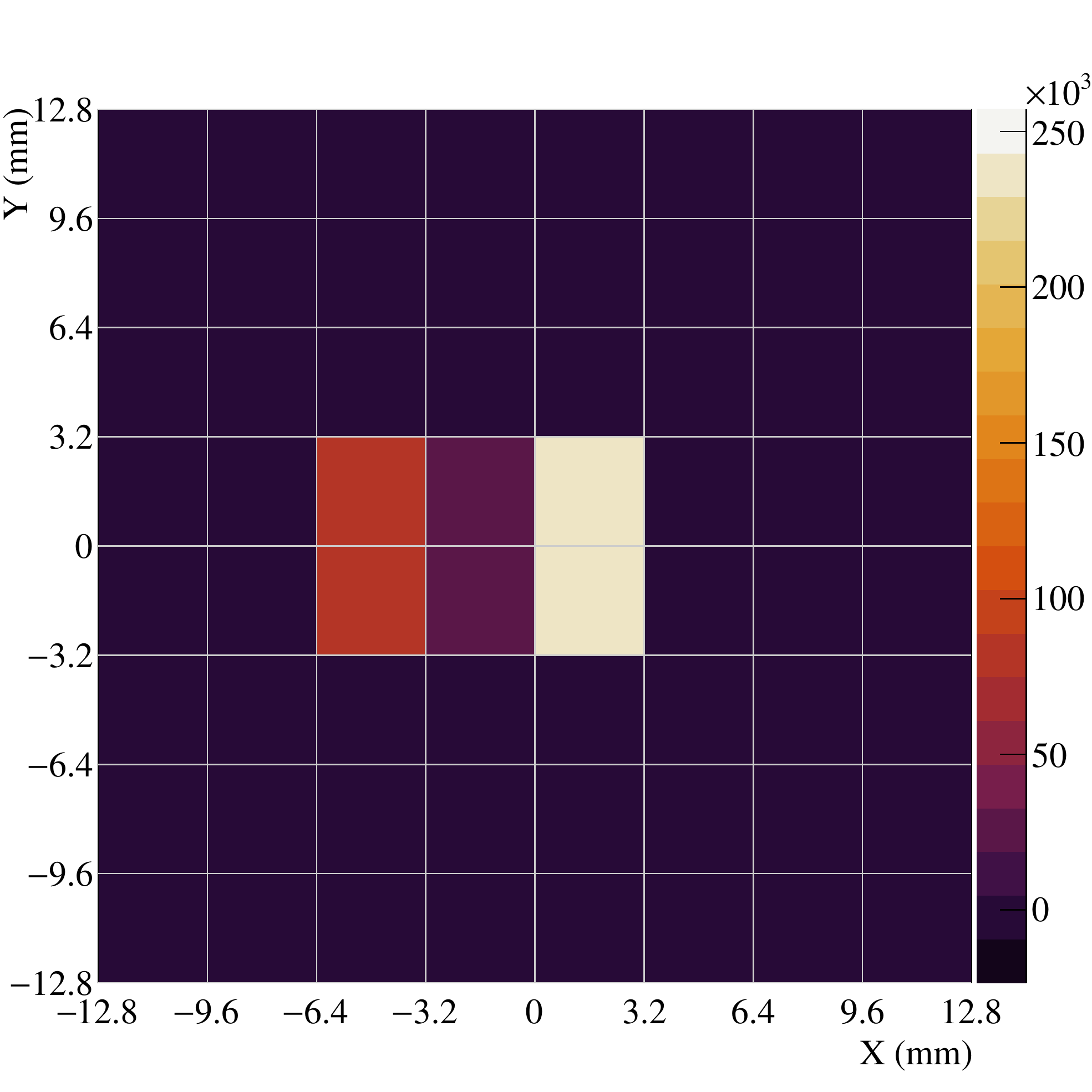}
  }
  \subfigure[Count map (70~deg)]{%
    \label{fig:map_465nm_70deg}
    \includegraphics[clip, width=6cm]{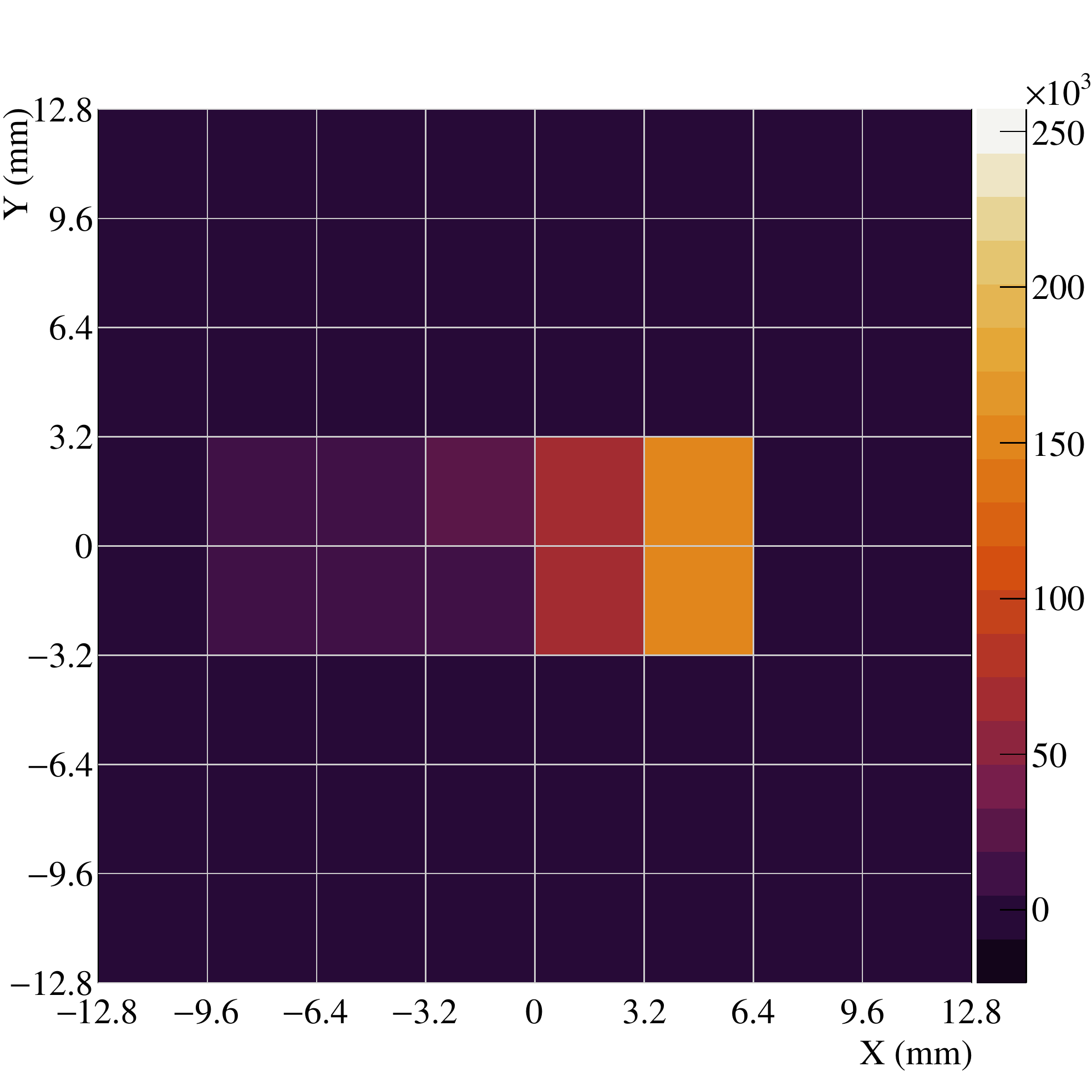}
  }\\
  \subfigure[Photon tracks (50~deg)]{%
    \label{fig:track50deg}
    \includegraphics[clip, width=6cm]{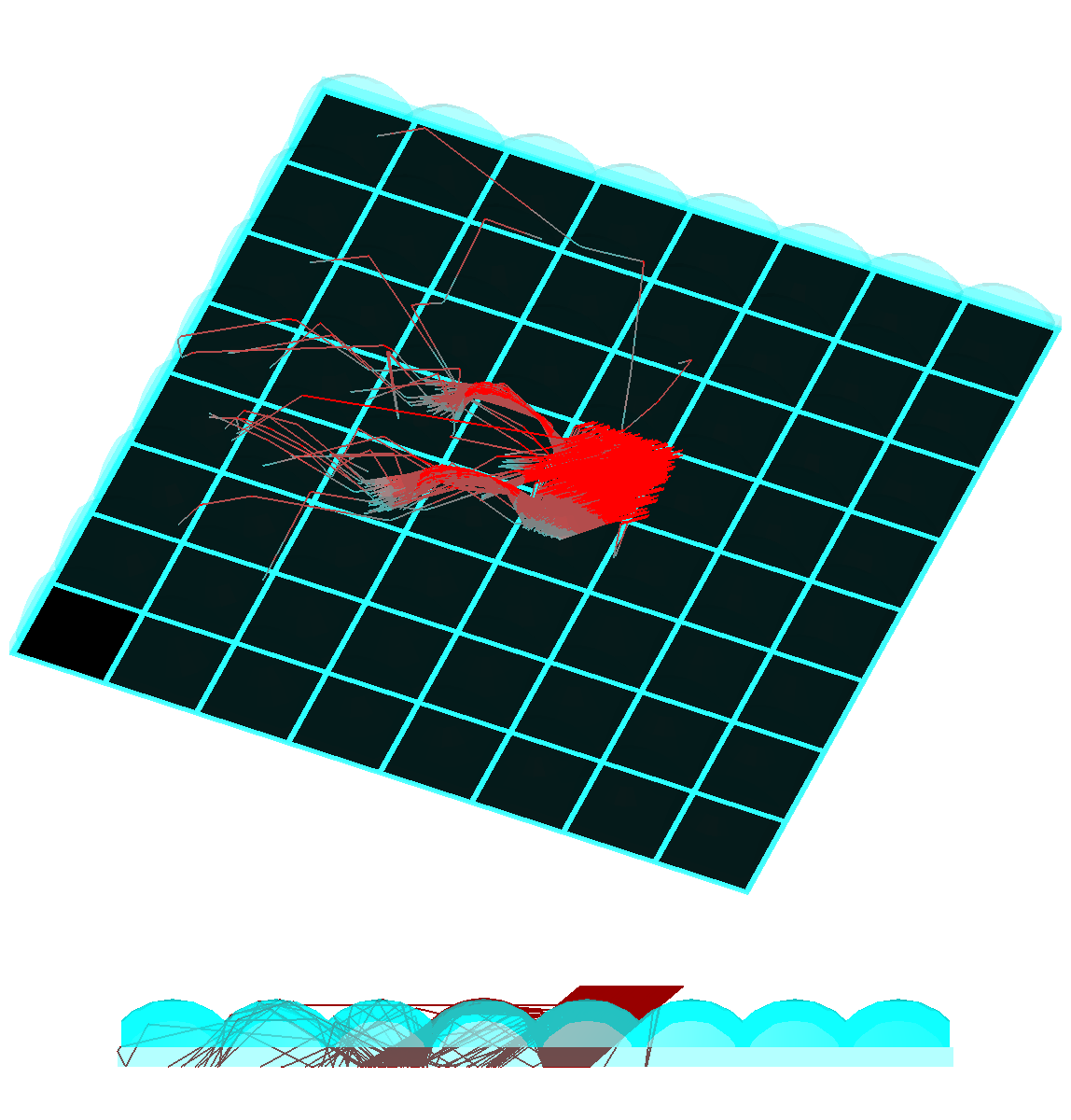}
  }
  \subfigure[Photon tracks (70~deg)]{%
    \label{fig:track70deg}
    \includegraphics[clip, width=6cm]{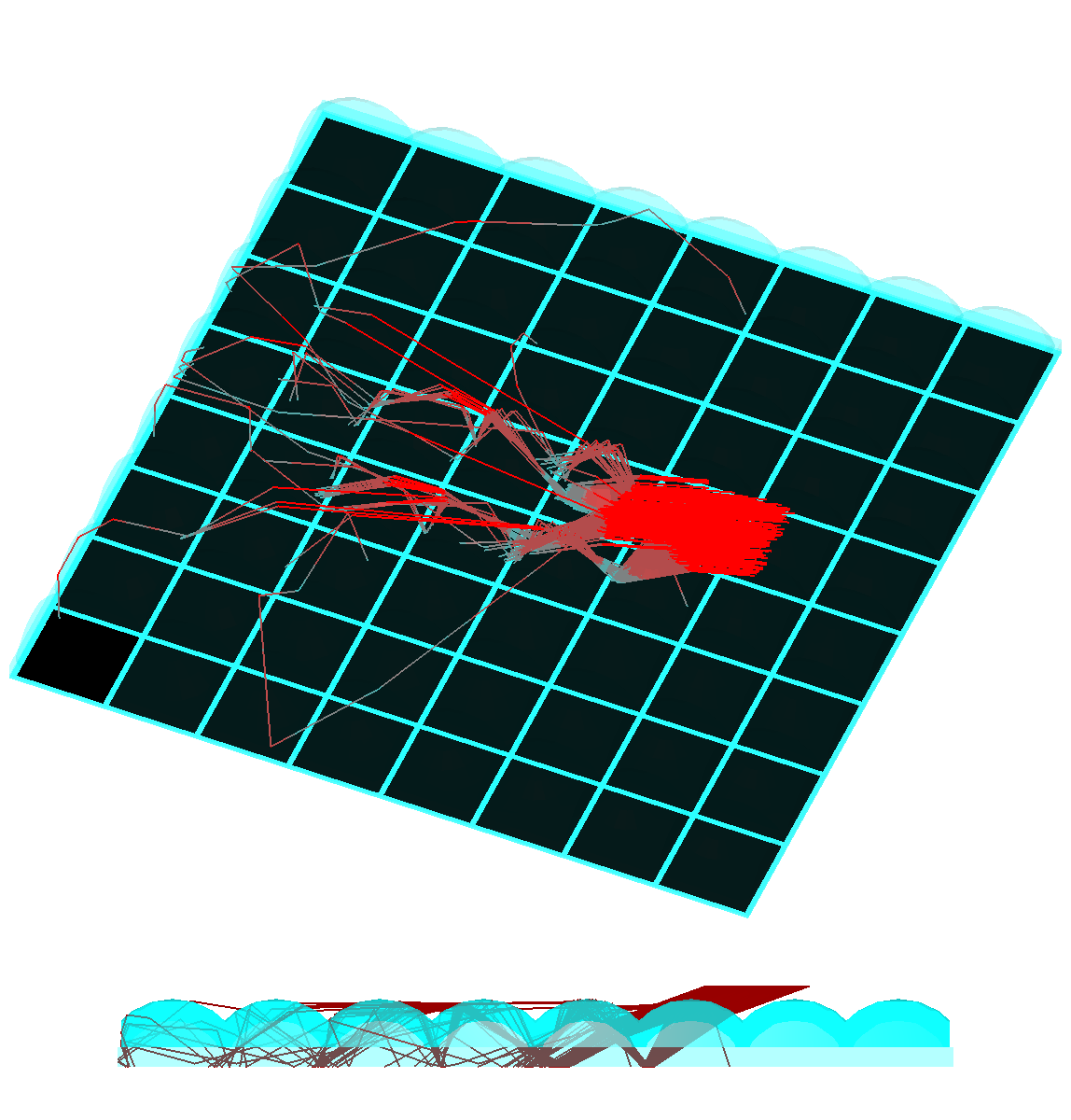}
  }
  \label{fig:map_track}
  \caption{(a) A simulated count map of detected photons in individual SiPM channels for 465-nm photons with an angle of incidence of 50~deg. (b) Same as (a) but for 70~deg. (c) Bird's eye view (top) and  side view (bottom) of the simulated photon tracks of an angle of incidence of 50~deg. The latter clearly shows the tracks of photons that were reflected on the silicon surface and then totally reflected on the lens array surface. (d) Same as (c) but for 70~deg.}
\end{figure}

Simulated count maps of detected photons and their tracks for 50 and 70~deg are shown in Fig.~\ref{fig:map_track}. Here, due to the total reflection, incident photons beamed toward the central four pixels were detected by not only these pixels but also the neighbor ones. In the case of 50~deg, photons were focused onto the two left neighbor pixels. In the case of 70~deg, a part of the beamed photons first entered the lenses of the two right neighbor pixels and the refracted tracks reached the silicone surface of the same pixels. These complex reflection and refraction processes resulted in a non-flat PDE increase at 40--70~deg.

\section{Measurement}
\label{sec:measurement}

We measured the actual performance of the lens array via uniform illumination of parallel LED beam flashes on the SiPM surface. Comparing the numbers of detected photons by one of the 64 channels of the SiPM array with the lens array being coupled and removed, we calculated the relative PDE increase from their ratio for various angles of incidence. Fig.~\ref{fig:sim_meas} shows the measurement results from $-75$ to $+75$~deg. Note that a uniform $3.2\times3.2$-$\mathrm{mm}^2$ beam was assumed in the ROBAST simulation, although parallel LED flashes were illuminated on the whole area of the SiPM array in our measurement. Thus, in Fig.~\ref{fig:sim_meas}, we can only provide a direct comparison between ``Sim. (Total),'' ``Measured (Pos. Rot.)'' (0--$+75$~deg), and ``Measured (Neg. Rot.)'' ($-75$--0~deg).

PDE increase measured in the all LED colors was roughly 10\%--20\% in 0--40~deg, which is consistent with the simulation result. Nonetheless, the expected large relative increase of more than 40\% in 50--70~deg was not necessarily as high as the simulation, whereas there was the occurrence of two peak structures at approximately 52 and 63~deg in the measurement. This result implied that reflectance at the media boundary between the silicon substrate and the protection silicone resin was significantly smaller than a simple assumption of Fresnel reflection. Moreover, the thin $\mathrm{SiO}_2$ and $\mathrm{Si}_3\mathrm{N}_4$ layers existing between these two media were expected to work as antireflection coating to improve the PDE; on the contrary, the quantum optical property of the layers was not considered in the simulation.

\section{Conclusion}

Through a ray-tracing simulation and a measurement, we showed the idea of coupling a spherical lens array on a SiPM array to increase their effective active area or to recover their physical 0.2-mm pixel gap. Based on our measurements, the relative PDE increase in the target angle regions of 30--60~deg was approximately 10\%--30\% for 310, 375, 465, and 635~nm of LED colors. Nonetheless, the simulation did not reproduce the measurement, particularly for large angles of incidence, probably due to imperfection in our optical treatment of the thin multi layers. This could be resolved by better modeling of the layers for further assessment and lens shape optimization. Degradation of focused image on the SiPM array due to the photon propagation onto the neighboring pixels as well as possible increase of optical crosstalk\cite{cite:OCT} should also be evaluated.

\end{document}